\documentclass[11pt,superscriptaddress,aps,prd,preprint,showpacs]{revtex4}

\usepackage[dvips]{graphicx}
\usepackage{amsfonts}
\usepackage{slashed}
\usepackage[latin1]{inputenc}
\usepackage{amsmath}
\usepackage{hyperref}
\usepackage[svgnames]{xcolor}

\begin{document}

\title{On perturbative aspects of a nonminimal Lorentz-violating QED with CPT-odd dimension-5 terms}

\author{T. Mariz}
\affiliation{Instituto de F\'{\i}sica, Universidade Federal de Alagoas,\\ 57072-900, Macei\'o, Alagoas, Brazil}
\email{tmariz,rmartinez@fis.ufal.br}

\author{R. Martinez}
\affiliation{Instituto de F\'{\i}sica, Universidade Federal de Alagoas,\\ 57072-900, Macei\'o, Alagoas, Brazil}
\email{tmariz,rmartinez@fis.ufal.br}

\author{J. R. Nascimento}
\affiliation{Departamento de F\'{\i}sica, Universidade Federal da Para\'{\i}ba,\\
 Caixa Postal 5008, 58051-970, Jo\~ao Pessoa, Para\'{\i}ba, Brazil}
\email{jroberto,petrov@fisica.ufpb.br}

\author{A. Yu. Petrov}
\affiliation{Departamento de F\'{\i}sica, Universidade Federal da Para\'{\i}ba,\\
 Caixa Postal 5008, 58051-970, Jo\~ao Pessoa, Para\'{\i}ba, Brazil}
\email{jroberto,petrov@fisica.ufpb.br}

\begin{abstract}
We consider the Lorentz-violating extended QED involving all nonminimal dimension-5 additive CPT-odd terms. For this theory, we investigate the generation of the Carroll-Field-Jackiw (CFJ) term and its higher-derivative counterparts of the first order in any of these nonminimal couplings. The CFJ term is demonstrated to vanish in the dimensional regularization scheme. We also study the question of higher-derivative divergent contributions and demonstrate that they can be eliminated by considering a given proportionality between the coefficients.
\end{abstract}


\maketitle

\section{Introduction}

The Lorentz-violating (LV) modifications of various field theory models are introduced through adding new terms proportional to constant vectors or, in general, tensors \cite{Colladay:1996iz,Colladay:1998fq,Kostelecky:2003fs}. Many examples of such additive terms have been proposed, giving origin both to minimal extensions, which involve only operators of dimensions up to 4, so, they include neither higher derivatives nor non-renormalizable couplings \cite{KosPic}, and the nonminimal ones where higher-dimension operators listed in \cite{Ding:2016lwt,KosLi,Kostelecky:2020hbb} are introduced. Among various aspects of new LV theories, their possible perturbative impacts are of special interest justified by the fact that perturbative generation of first known LV term, that is, the Carroll-Field-Jackiw (CFJ) term \cite{CFJ}, performed in \cite{JK}, implied formulating a whole methodology for inducing new LV additive terms in the gauge sector. 

According to this methodology, one starts with an extended spinor QED involving additive CPT-odd dimension-5 operators. As a result, new LV terms for the gauge field are generated when we consider one-loop corrections. Besides the CFJ term, also other additive terms in the gauge sector have been generated in this manner. One can mention, e.g., the CPT-even aether term \cite{aether,aether0,aether1,Mariz:2016ooa}, and the higher-derivative LV terms, namely, the Myers-Pospelov and the higher-derivative CFJ-like terms \cite{MNP,TMHD}, with all additive LV couplings being the minimal ones, except for the magnetic coupling originally introduced in \cite{magn}. An important feature of all these results consists in the fact that they are all either finite or renormalizable. Nevertheless, it should be noted that in principle, these terms can arise even if the magnetic coupling is not used, see \cite{aether1,MNP}. Further, perturbative impacts of the CPT-even nonminimal coupling $\kappa^{\alpha\beta\mu\nu}\bar{\psi}\sigma_{\mu\nu}F_{\alpha\beta}\psi$ were studied in \cite{Maluf,Carvalho:2018vtr}.

However, it is clear that the quantum impacts of nonminimal couplings naturally need further studies. While a list of all possible LV vertices with dimensions up to 6 is presented in \cite{Ding:2016lwt,KosLi}, it was shown in \cite{ourrev} that, although tree-level effects generated by many of these dimensions-5 and 6 vertices have been intensively studied, see, e.g., \cite{Casana:2012vu,Casana:2012yj,Araujo:2019txk,Ferreira:2019lpu}, only a few of such vertices have been really treated within the perturbative context up to now. The present study is aimed to follow this line, explicitly, to study the perturbative corrections in a nonminimal LV extended QED involving all CPT-odd dimension-5 operators proposed in \cite{KosLi}.

The structure of the paper looks like follows. In section \ref{model}, we define our model and calculate the effective action.  We perform the one-loop calculations in sections \ref{de} and \ref{fp} to study the possible generation of the CFJ term and the cancelation of the divergences that can appear in higher-derivative terms. Finally, in section \ref{summ}, we discuss our results.

\section{Nonminimally extended Lorentz-violating QED}\label{model}

In this paper, we are interested in analyzing the following nonminimal LV extended QED Lagrangian:
\begin{eqnarray}\label{start}
\mathcal{L}_{\psi} &=& \bar{\psi}(i\slashed{D} -m)\psi -\textstyle{1\over2}m^{(5)\alpha\beta}\bar{\psi}iD_{(\alpha}iD_{\beta)}\psi +\mathrm{h.c.} -\textstyle{1\over2}im_5^{(5)\alpha\beta}\bar{\psi}\gamma_5iD_{(\alpha}iD_{\beta)}\psi +\mathrm{h.c.} \nonumber\\
&&-\textstyle{1\over2}a^{(5)\mu \alpha \beta}\bar{\psi}\gamma_{\mu}iD_{(\alpha}iD_{\beta)}\psi +\mathrm{h.c.} -\textstyle{1\over2}b^{(5)\mu \alpha \beta}\bar{\psi}\gamma_5 \gamma_{\mu}iD_{(\alpha}iD_{\beta)}\psi +\mathrm{h.c.} \nonumber\\
&&-\textstyle{1\over4}H^{(5)\mu\nu\alpha\beta}\bar{\psi}\sigma_{\mu\nu}iD_{(\alpha}iD_{\beta)}\psi +\mathrm{h.c.} -\textstyle{1\over2}m^{(5)\alpha\beta}_F\bar{\psi}F_{\alpha\beta}\psi -\textstyle{1\over2}im_{5F}^{(5)\alpha\beta}\bar{\psi}\gamma_5F_{\alpha\beta}\psi \nonumber\\
&&-\textstyle{1\over2}a_F^{(5)\mu \alpha \beta} \bar{\psi}\gamma_\mu F_{\alpha \beta} \psi -\textstyle{1\over2}b_F^{(5)\mu \alpha \beta} \bar{\psi}\gamma_5 \gamma_\mu F_{\alpha \beta} \psi -\textstyle{1\over4}H^{(5)\mu\nu\alpha\beta}_F\bar{\psi}\sigma_{\mu\nu}F_{\alpha\beta}\psi,
\end{eqnarray}
where $F_{\alpha \beta} = \partial_\alpha A_\beta - \partial_\beta A_\alpha $, $D_\mu\psi= \partial_\mu\psi+ie A_\mu\psi$, and
\begin{eqnarray}
iD_{(\alpha} i D_{\beta)}\psi&=&\textstyle{1\over2}\left(iD_\alpha i D_\beta + iD_\beta iD_\alpha\right)\psi\nonumber \\
&=&-\partial_\alpha \partial_\beta\psi -ie[A_\beta \partial_\alpha+A_\alpha \partial_\beta +\textstyle{1\over2}(\partial_\alpha A_\beta + \partial_\beta A_\alpha)]\psi+e^2A_\alpha A_\beta\psi.
\end{eqnarray}
The expression (\ref{start}) includes all dimension-5 LV couplings defined in \cite{KosLi}. In fact, our aim will consist of studying the CPT-odd contributions to the one-loop effective action of the gauge field, e.g., for the one-derivative contribution, the CFJ term. It is clear that exclusively the terms involving odd-rank constant tensors must be considered since only contractions of such tensors with the Minkowski metric and the Levi-Civita symbol, in four-dimensional space-time, could generate a constant axial vector necessary for forming the CFJ term (and its higher-derivative counterpart \cite{Leite:2013pca}). Therefore, within this study, we deal with all possible CPT-odd dimension-5 couplings. We note that, as we will see further, such couplings can generate higher-derivative terms in the pure gauge sector as well.
This allows to reduce our Lagrangian to
\begin{eqnarray}\label{1}
\mathcal{L}_{\psi} &=& \bar{\psi}(i\slashed{D} -m)\psi -\textstyle{1\over2}a^{(5)\mu \alpha \beta}\bar{\psi}\gamma_{\mu}iD_{(\alpha}iD_{\beta)}\psi +\mathrm{h.c.} -\textstyle{1\over2}b^{(5)\mu \alpha \beta}\bar{\psi}\gamma_5 \gamma_{\mu}iD_{(\alpha}iD_{\beta)}\psi +\mathrm{h.c.}\nonumber\\
&&-\textstyle{1\over2}a_F^{(5)\mu \alpha \beta} \bar{\psi}\gamma_\mu F_{\alpha \beta} \psi -\textstyle{1\over2}b_F^{(5)\mu \alpha \beta} \bar{\psi}\gamma_5 \gamma_\mu F_{\alpha \beta} \psi.
\end{eqnarray}
We can as well rewrite the expression (\ref{1}) as follows:
\begin{eqnarray}
\mathcal{L}_{\psi} &=& \bar{\psi} [i\slashed{\partial} +(a^{(5)\mu \alpha \beta}+b^{(5)\mu \alpha \beta} \gamma_5) \gamma_{\mu}\partial_\alpha \partial_\beta-m -e\slashed{A} +ie(a^{(5)\mu \alpha \beta}+b^{(5)\mu \alpha \beta} \gamma_5)\gamma_\mu \nabla_\alpha A_\beta \nonumber\\ 
&& -2e(a_F^{(5)\mu \alpha \beta}+b_F^{(5)\mu \alpha \beta} \gamma_5) \gamma_\mu \partial_\alpha A_\beta -e^2(a^{(5)\mu \alpha \beta}+b^{(5)\mu \alpha \beta} \gamma_5) \gamma_{\mu}A_\alpha A_\beta]\psi,
\end{eqnarray}
where we introduced the definition $\nabla_\alpha A_\beta\equiv 2A_\beta\partial_\alpha +(\partial_\alpha A_\beta)$.

The corresponding fermionic generating functional is
\begin{equation}
Z = \int D\bar\psi D\psi e^{i\int d^4x{\cal L}_\psi} = e^{iS_\mathrm{eff}},
\end{equation}
so that, by integrating out the spinor fields, we obtain the one-loop effective action  of the gauge field 
\begin{eqnarray}\label{Seff}
S_\mathrm{eff} &=& -i\mathrm{Tr}\ln[\slashed{p} -(a^{(5)\mu \alpha \beta}+b^{(5)\mu \alpha \beta} \gamma_5) \gamma_{\mu}p_\alpha p_\beta-m -e\slashed{A} +e(a^{(5)\mu \alpha \beta}+b^{(5)\mu \alpha \beta} \gamma_5)\gamma_\mu \\ 
&& \times\nabla_\alpha(p,k) A_\beta +2ie(a_F^{(5)\mu \alpha \beta}+b_F^{(5)\mu \alpha \beta} \gamma_5) \gamma_\mu k_\alpha A_\beta -e^2(a^{(5)\mu \alpha \beta}+b^{(5)\mu \alpha \beta} \gamma_5) \gamma_{\mu}A_\alpha A_\beta]\psi, \nonumber
\end{eqnarray}
with, in the momentum space, $\nabla_\alpha(p,k)=2p_\alpha+k_\alpha$, $i\partial_\alpha \psi=p_\alpha \psi$, and $i\partial_\alpha A_\beta =k_\alpha A_\beta$. Here, $\mathrm{Tr}$ stands for the trace over the Dirac matrices, as well as the trace over the integration in momentum and coordinate spaces.

We can expand Eq.~(\ref{Seff}) in power series in external fields as 
\begin{equation}
S_\mathrm{eff}=S_\mathrm{eff}^{(0)}+\sum_{n=1}^\infty S_\mathrm{eff}^{(n)},
\end{equation} 
where $S_\mathrm{eff}^{(0)}=-i\mathrm{Tr}\ln G^{-1}(p)$ and 
\begin{eqnarray}
S_\mathrm{eff}^{(n)} &=& \frac{i}{n}\mathrm{Tr}\{G(p)[e\slashed{A} -e(a^{(5)\mu \alpha \beta}+b^{(5)\mu \alpha \beta} \gamma_5)\gamma_\mu \nabla_\alpha(p,k) A_\beta \nonumber\\
&& -2ie(a_F^{(5)\mu \alpha \beta}+b_F^{(5)\mu \alpha \beta} \gamma_5) \gamma_\mu k_\alpha A_\beta +e^2(a^{(5)\mu \alpha \beta}+b^{(5)\mu \alpha \beta} \gamma_5) \gamma_{\mu}A_\alpha A_\beta]\}^n,
\end{eqnarray}
with 
\begin{equation}
G(p)=\frac1{\slashed{p} -(a^{(5)\mu \alpha \beta}+b^{(5)\mu \alpha \beta} \gamma_5) \gamma_{\mu}p_\alpha p_\beta -m}.
\end{equation}
Since, for this step, we are interested in the induced CPT-odd terms, we need to work only with terms of first order in $a^{(5)\mu \alpha \beta}$, $b^{(5)\mu \alpha \beta}$, $a_F^{(5)\mu \alpha \beta}$, $b_F^{(5)\mu \alpha \beta}$, and second order in $A^\mu$. After evaluating the trace over the coordinate space, by using the key identity of the derivative expansion method $A_\mu(x)G(p)=G(p-k)A_\mu(x)$ \cite{DerEx} and integrating over momenta, we can write two lower contributions to the one-loop result for the quadratic action $A_\mu$ as
\begin{equation}
S^{(1)}_\mathrm{eff}=i\int d^4x \Pi_1^{\mu \nu}A_\mu A_\nu,
\end{equation} with
\begin{equation}\label{Pi1}
\Pi^{\mu\nu}_1= e^2\int \frac{d^4p}{(2\pi)^4}\mathrm{tr}\,G(p)(a^{(5)\lambda\mu\nu}+b^{(5)\lambda\mu\nu} \gamma_5) \gamma_{\lambda},
\end{equation}
and
\begin{equation}\label{CS2}
S^{(2)}_\mathrm{eff}=\frac{i}{2}\int d^4x (\Pi^{\mu\nu}_2+\Pi^{\mu\nu}_3+\Pi^{\mu\nu}_4+\Pi^{\mu\nu}_5+\Pi^{\mu\nu}_6)A_\mu A_\nu,
\end{equation}
with
\begin{subequations}\label{Pi1-6} 
\begin{eqnarray}
\Pi^{\mu \nu}_2&=&-e^2 \int \frac{d^4p}{(2\pi)^4}\mathrm{tr}\,G(p) (a^{(5)\kappa\lambda\mu}+b^{(5)\kappa\lambda\mu} \gamma_5)\gamma_\kappa \nabla_\lambda(p,k)G(p-k) \gamma^\nu,\\
\Pi^{\mu \nu}_3&=&-e^2\int \frac{d^4p}{(2\pi)^4}\mathrm{tr}\,G(p) \gamma^\mu G(p-k)(a^{(5)\kappa\lambda\nu}+b^{(5)\kappa\lambda\nu} \gamma_5)\gamma_\kappa \nabla_\lambda(p-k,-k),\\
\Pi^{\mu \nu}_4&=&e^2\int \frac{d^4p}{(2\pi)^4}\mathrm{tr}\,G(p) \gamma^\mu G(p-k)\gamma^\nu, \\
\Pi^{\mu \nu}_5&=&-2ie^2\int\frac{d^4p}{(2\pi)^4}\mathrm{tr}\, G(p)(a_F^{(5)\kappa\lambda\mu}+b_F^{(5)\kappa\lambda\mu}\gamma_5) \gamma_\kappa k_\lambda G(p-k)\gamma^\nu,\\
\Pi^{\mu \nu}_6&=&-2ie^2\int \frac{d^4p}{(2\pi)^4}\mathrm{tr}\, G(p)\gamma^\mu G(p-k)(a_F^{(5)\kappa\lambda\nu}+b_F^{(5)\kappa\lambda\nu}\gamma_5) \gamma_\kappa (-k_\lambda).
\end{eqnarray}
\end{subequations}
Let us now single out the contributions of the first order in the coefficients for Lorentz violation. For this, we must first take into account the expansion of the propagator $G(p)$, given by
\begin{equation}
G(p)=S(p)+S(p) (a^{(5)\mu \alpha \beta}+b^{(5)\mu \alpha \beta} \gamma_5) \gamma_{\mu}p_\alpha p_\beta S(p)+\cdots,
\end{equation}
with $S(p)=(\slashed{p}+m)^{-1}$. Then, we can rewrite the above expressions as follows:
\begin{subequations}\label{foc}
\begin{eqnarray}
\Pi^{\mu\nu}_1&=& e^2\int \frac{d^4p}{(2\pi)^4}\mathrm{tr}\,S(p)(a^{(5)\lambda\mu\nu}+b^{(5)\lambda\mu\nu} \gamma_5) \gamma_{\lambda}, \\
\Pi^{\mu \nu}_2&=&-e^2 \int \frac{d^4p}{(2\pi)^4}\mathrm{tr}\,S(p) (a^{(5)\alpha\kappa\mu}+b^{(5)\alpha\kappa\mu} \gamma_5)\gamma_\alpha \nabla_\kappa(p,k)S(p-k) \gamma^\nu,\\
\Pi^{\mu \nu}_3&=&-e^2\int \frac{d^4p}{(2\pi)^4}\mathrm{tr}\,S(p) \gamma^\mu S(p-k)(a^{(5)\alpha\kappa\nu}+b^{(5)\alpha\kappa\nu} \gamma_5)\gamma_\alpha \nabla_\kappa(p-k,-k),\\
\Pi^{\mu \nu}_4&=&e^2\int \frac{d^4p}{(2\pi)^4}\mathrm{tr} \left[S(p) (a^{(5)\alpha\kappa\lambda}+b^{(5)\alpha\kappa\lambda}\gamma_5)\gamma_{\alpha}p_\kappa p_\lambda S(p) \gamma^\mu S(p-k)\gamma^\nu \right. \nonumber\\
&&\left.+S(p) \gamma^\mu S(p-k) (a^{(5)\alpha\kappa\lambda}+b^{(5)\alpha\kappa\lambda}\gamma_5)\gamma_{\alpha}(p-k)_\kappa (p-k)_\lambda S(p-k)\gamma^\nu\right], \\
\Pi^{\mu \nu}_5&=&-2ie^2\int\frac{d^4p}{(2\pi)^4}\mathrm{tr}\, S(p)(a_F^{(5)\alpha\kappa\mu}+b_F^{(5)\alpha\kappa\mu}\gamma_5) \gamma_\alpha k_\kappa S(p-k)\gamma^\nu,\\
\Pi^{\mu \nu}_6&=&-2ie^2\int \frac{d^4p}{(2\pi)^4}\mathrm{tr}\, S(p)\gamma^\mu S(p-k)(a_F^{(5)\alpha\kappa\nu}+b_F^{(5)\alpha\kappa\nu}\gamma_5) \gamma_\alpha (-k_\kappa).
\end{eqnarray}
\end{subequations}
It is easy to see, trivially, that $\Pi_1^{\mu\nu}$ vanishes. 

In the next sections, let us analyze the questions of generating the CFJ term and potential divergences that can arise in higher-derivative terms.

\section{Derivative expansion}\label{de}

This section aims to obtain the low-energy effective action in our theory, i.e., the CFJ action. To do it, we employ the derivative expansion framework \cite{DerEx} and keep only the one-derivative term which is sufficient for our purposes. We note that to obtain the CFJ term, we must have the Levi-Civita symbol contracted to an axial vector. Both $a^{(5)\mu\alpha\beta}$ and $b^{(5)\mu\alpha\beta}_F$ cannot yield such a structure. Indeed, neither $a^{(5)\mu\alpha\beta}$ nor $b^{(5)\mu\alpha\beta}_F$ can be represented in the form of a constant axial vector multiplied, 
 by some invariant tensor, either the Levi-Civita symbol (for $a^{(5)\mu\alpha\beta}$) or Minkowski metric (for $b^{(5)\mu\alpha\beta}_F$), in the manner allowing to yield the CFJ term. So, we rest with couplings proportional to $b^{(5)\alpha\kappa\mu}$ and $a_F^{(5)\alpha\kappa\mu}$. 

Then, using the expansion
\begin{equation}
S(p-k)=S(p)+S(p)\slashed{k}S(p)+\cdots
\end{equation}
in the expressions (\ref{foc}), we get $\Pi^{\mu\nu}_2\to\Pi^{\mu\nu}_\mathrm{CFJ2}=\Pi^{\mu\nu}_{2,1}+\Pi^{\mu\nu}_{2,2}$, where
\begin{subequations}
\begin{eqnarray}
\Pi^{\mu \nu}_{2,1}&=&-e^2 \int \frac{d^4p}{(2\pi)^4}\mathrm{tr}\,S(p) b^{(5)\alpha\kappa\mu} \gamma_5 \gamma_\alpha \nabla_\kappa(p,k)S(p) \gamma^\nu, \\
\Pi^{\mu \nu}_{2,2}&=&-e^2 \int \frac{d^4p}{(2\pi)^4}\mathrm{tr}\,S(p) b^{(5)\alpha\kappa\mu} \gamma_5 \gamma_\alpha \nabla_\kappa(p,0)S(p)\slashed{k}S(p) \gamma^\nu,
\end{eqnarray}
\end{subequations}
as well as $\Pi^{\mu\nu}_3\to\Pi^{\mu\nu}_\mathrm{CFJ3}=\Pi^{\mu\nu}_{3,1}+\Pi^{\mu\nu}_{3,2}$, with
\begin{subequations}
\begin{eqnarray}
\Pi^{\mu \nu}_{3,1}&=&-e^2\int \frac{d^4p}{(2\pi)^4}\mathrm{tr}\,S(p) \gamma^\mu S(p) b^{(5)\alpha\kappa\nu} \gamma_5 \gamma_\alpha \nabla_\kappa(p-k,-k), \\
\Pi^{\mu \nu}_{3,2}&=&-e^2\int \frac{d^4p}{(2\pi)^4}\mathrm{tr}\,S(p) \gamma^\mu S(p)\slashed{k}S(p)b^{(5)\alpha\kappa\nu} \gamma_5 \gamma_\alpha \nabla_\kappa(p,0),
\end{eqnarray}
\end{subequations}
and $\Pi^{\mu\nu}_4\to\Pi^{\mu\nu}_\mathrm{CFJ4}=\Pi^{\mu\nu}_{4,1}+\Pi^{\mu\nu}_{4,2}+\Pi^{\mu\nu}_{4,3}+\Pi^{\mu\nu}_{4,4}+\Pi^{\mu\nu}_{4,5}$, with
\begin{subequations}
\begin{eqnarray} 
\Pi^{\mu \nu}_{4,1}&=&e^2\int \frac{d^4p}{(2\pi)^4}\mathrm{tr} S(p) b^{(5)\alpha\kappa\lambda}\gamma_5 \gamma_{\alpha}p_\kappa p_\lambda S(p) \gamma^\mu S(p)\slashed{k}S(p)\gamma^\nu, \\
\Pi^{\mu \nu}_{4,2}&=&e^2\int \frac{d^4p}{(2\pi)^4}\mathrm{tr} S(p) \gamma^\mu S(p)\slashed{k}S(p) b^{(5)\alpha\kappa\lambda}\gamma_5 \gamma_{\alpha}p_\kappa p_\lambda S(p)\gamma^\nu, \\
\Pi^{\mu \nu}_{4,3}&=&e^2\int \frac{d^4p}{(2\pi)^4}\mathrm{tr} S(p) \gamma^\mu S(p) b^{(5)\alpha\kappa\lambda}\gamma_5 \gamma_{\alpha}(-k_\kappa) p_\lambda S(p)\gamma^\nu, \\
\Pi^{\mu \nu}_{4,4}&=&e^2\int \frac{d^4p}{(2\pi)^4}\mathrm{tr} S(p) \gamma^\mu S(p) b^{(5)\alpha\kappa\lambda}\gamma_5 \gamma_{\alpha}p_\kappa (-k_\lambda) S(p)\gamma^\nu, \\
\Pi^{\mu \nu}_{4,5}&=&e^2\int \frac{d^4p}{(2\pi)^4}\mathrm{tr} S(p) \gamma^\mu S(p) b^{(5)\alpha\kappa\lambda}\gamma_5 \gamma_{\alpha}p_\kappa p_\lambda S(p)\slashed{k}S(p)\gamma^\nu, 
\end{eqnarray}
\end{subequations}
and finally, $\Pi^{\mu\nu}_5\to\Pi^{\mu\nu}_\mathrm{CFJ5}=\Pi^{\mu\nu}_{5,1}$ and $\Pi^{\mu\nu}_6\to\Pi^{\mu\nu}_\mathrm{CFJ6}=\Pi^{\mu\nu}_{6,1}$, where
\begin{subequations}
\begin{eqnarray} 
\Pi^{\mu \nu}_{5,1} &=&-2ie^2\int\frac{d^4p}{(2\pi)^4}\mathrm{tr}\, S(p) a_F^{(5)\alpha\kappa\mu} \gamma_\alpha k_\kappa S(p)\gamma^\nu, \\
\Pi^{\mu \nu}_{6,1}&=&-2ie^2\int \frac{d^4p}{(2\pi)^4}\mathrm{tr}\, S(p)\gamma^\mu S(p) a_F^{(5)\alpha\kappa\nu} \gamma_\alpha (-k_\kappa).
\end{eqnarray}
\end{subequations}

Calculating the trace, considering the replacement $d^4p/(2\pi)^4 \to \mu^{4-D}d^Dp/(2\pi)^D$, and using the identities
\begin{subequations}\label{Rpp}
\begin{eqnarray}
\int \frac{d^Dp}{(2\pi)^D}p_\mu p_\nu f(p^2) &=& \frac{g_{\mu\nu}}{D}\int \frac{d^Dp}{(2\pi)^D}p^2 f(p^2), \\
\int \frac{d^Dp}{(2\pi)^D}p_\kappa p_\lambda p_\mu p_\nu f(p^2) &=& \frac{g_{\kappa\lambda}g_{\mu\nu}+g_{\kappa\mu}g_{\lambda\nu}+g_{\kappa\nu}g_{\lambda\mu}}{D(D+2)}\int \frac{d^Dp}{(2\pi)^D}p^4 f(p^2),
\end{eqnarray}
\end{subequations}
we find that the relevant contributions are
\begin{subequations}\label{E0}
\begin{eqnarray}
\Pi^{\mu \nu}_\mathrm{CFJ2} &=& \frac{8ie^2}{D} \mu^{4-D} \int \frac{d^Dp}{(2\pi)^D}\frac{p^2}{(p^2-m^2)^2} b^{(5)\alpha\kappa\lambda}  {g_\lambda}^\mu  k^{\beta} {\epsilon_{\alpha\beta\kappa}}^{\nu},\\
\Pi^{\mu \nu}_\mathrm{CFJ3} &=& -\frac{8ie^2}{D} \mu^{4-D} \int \frac{d^Dp}{(2\pi)^D}\frac{p^2}{(p^2-m^2)^2} b^{(5)\alpha\kappa\lambda} {g_\lambda}^\nu  k^\beta {\epsilon_{\alpha\beta\kappa}}^\mu,\\
\Pi^{\mu \nu}_\mathrm{CFJ4} &=& \frac{4ie^2}{D} \mu^{4-D} \int\frac{d^Dp}{(2\pi)^D}\frac{p^2}{(p^2-m^2)^2}b^{(5)\alpha \kappa \lambda}\left(3g_{\lambda\kappa}k^\beta {\epsilon^{\mu \nu}}_{ \alpha \beta}+k_\kappa{\epsilon^{\mu \nu}}_{ \alpha \lambda}+k_\lambda{\epsilon^{\mu \nu}}_{ \alpha \kappa}\right)\\ 
&& \nonumber -\frac{16ie^2}{D(D+2)} \mu^{4-D} \int\frac{d^Dp}{(2\pi)^D}\frac{p^4}{(p^2-m^2)^3}b^{(5)\alpha \kappa \lambda}\left(3g_{\lambda\kappa}k_\beta {\epsilon^{\mu \nu}}_{\alpha \beta} - {g_\kappa}^\mu k_\beta {\epsilon^{\nu \beta}}_{ \lambda \alpha} \right.\\&&\nonumber\left.-{g_\lambda}^\mu k_\beta {\epsilon^{\nu \beta}}_ {\kappa \alpha}+{g_\kappa}^\nu k_\beta{\epsilon^{\mu\beta }}_{\lambda\alpha}+{g_\lambda}^\nu k_\beta {\epsilon^{\mu \beta}}_{\kappa \alpha}+k_\kappa {\epsilon^{\mu \nu}}_{\alpha \lambda}+k_\lambda {\epsilon^{\mu \nu}}_{\alpha \kappa} \right)  , \\ 
\Pi^{\mu \nu}_\mathrm{CFJ5} &=& -\frac{8ie^2}{D} \mu^{4-D} \int\frac{d^Dp}{(2\pi)^D}\frac{1}{(p^2-m^2)^2} a_F^{(5)\alpha\kappa\lambda}   \left( 2p^2-D(p^2-m^2) \right) {g_\alpha}^\nu k_{\kappa }{g_\lambda}^\mu, \\ 
\Pi^{\mu \nu}_\mathrm{CFJ6} &=& \frac{8ie^2}{D} \mu^{4-D} \int\frac{d^Dp}{(2\pi)^D}\frac{1}{(p^2-m^2)^2} a_F^{(5)\alpha\kappa\lambda} \left( 2p^2-D(p^2-m^2)\right) {g_\alpha}^\mu k_\kappa {g_\lambda}^\nu,
\end{eqnarray}
\end{subequations}
with $\Pi^{\mu \nu}_\mathrm{CFJ}= \Pi^{\mu \nu}_\mathrm{CFJ2}+\Pi^{\mu \nu}_\mathrm{CFJ3}+\Pi^{\mu \nu}_\mathrm{CFJ4}+\Pi^{\mu \nu}_\mathrm{CFJ5}+\Pi^{\mu \nu}_\mathrm{CFJ6}$. These contributions will be evaluated one by one employing dimensional regularization. First, the tensors $\Pi^{\mu \nu}_\mathrm{CFJ2}$ and $\Pi^{\mu \nu}_\mathrm{CFJ3}$ turn out to be
\begin{eqnarray}
\Pi^{\mu \nu}_\mathrm{CFJ2}+\Pi^{\mu \nu}_\mathrm{CFJ3} &=& b^{(5)\alpha\kappa\lambda} 2^{2-D} \pi ^{-\frac{D}{2}} e^2 \mu ^{4-D} m^{D-2} \Gamma \left(1-\frac{D}{2}\right) \nonumber\\
&&\times k^\beta \left({g_\lambda}^\mu {\epsilon_{\alpha\beta\kappa}}^\nu -{g_\lambda}^\nu {\epsilon_{\alpha\beta\kappa}}^\mu\right),
\end{eqnarray}
where the gamma function $\Gamma \left(1-\frac{D}{2}\right)$ displays divergent behavior in $D=4$.
Next, the tensor $\Pi^{\mu \nu}_\mathrm{CFJ4}$ yields
\begin{eqnarray}
\Pi^{\mu \nu}_\mathrm{CFJ4} &=& -b^{(5)\alpha\kappa\lambda} 2^{1-D} \pi ^{-\frac{D}{2}} e^2 \mu ^{4-D} m^{D-2} \Gamma \left(1-\frac{D}{2}\right) \nonumber\\
&&\times k^\beta \left({g_\lambda}^\mu {\epsilon_{\alpha\beta\kappa}}^\nu +{g_\kappa}^\mu {\epsilon_{\alpha\beta\lambda}}^\nu -{g_\lambda}^\nu {\epsilon_{\alpha\beta\kappa}}^\mu -{g_\kappa}^\nu {\epsilon_{\alpha\beta\lambda}}^\mu\right).
\end{eqnarray}
Using the fact that $b^{(5)\alpha\kappa\lambda}$ is a symmetric tensor, i.e., $b^{(5)\alpha\lambda\kappa}=b^{(5)\alpha\kappa\lambda}$, we can rewrite $\Pi^{\mu\nu}_{CFJ4}$ as
\begin{eqnarray}
\Pi^{\mu \nu}_\mathrm{CFJ4} &=& -b^{(5)\alpha\kappa\lambda} 2^{2-D} \pi ^{-\frac{D}{2}} e^2 \mu ^{4-D} m^{D-2} \Gamma \left(1-\frac{D}{2}\right) \nonumber\\
&&\times k^\beta \left({g_\lambda}^\mu {\epsilon_{\alpha\beta\kappa}}^\nu -{g_\lambda}^\nu {\epsilon_{\alpha\beta\kappa}}^\mu\right).
\end{eqnarray}
The remaining contributions $\Pi^{\mu \nu}_\mathrm{CFJ5}$ and $\Pi^{\mu \nu}_\mathrm{CFJ6}$ vanish after the loop integration for any value of $D$, i.e., 
\begin{eqnarray}
\Pi^{\mu \nu}_\mathrm{CFJ5}+\Pi^{\mu \nu}_\mathrm{CFJ6}=0.
\end{eqnarray}
It is easy to see that the sum of all contributions is identically zero so that
\begin{eqnarray}\label{O0}
\Pi^{\mu \nu}_\mathrm{CFJ} = \Pi^{\mu \nu}_\mathrm{CFJ2}+\Pi^{\mu \nu}_\mathrm{CFJ3}+\Pi^{\mu \nu}_\mathrm{CFJ4}+\Pi^{\mu \nu}_\mathrm{CFJ5}+\Pi^{\mu \nu}_\mathrm{CFJ6} = 0.
\end{eqnarray}

At the same time, if we had used the symmetrizations $p_{\mu}p_{\nu}\to \frac14 g_{\mu\nu} p^2$ and $p_\kappa p_\lambda p_\mu p_\nu\to\frac1{24}(g_{\kappa\lambda}g_{\mu\nu}+g_{\kappa\mu}g_{\lambda\nu}+g_{\kappa\nu}g_{\lambda\mu})p^4$ instead of (\ref{Rpp}), we could have a non-zero result, given by 
\begin{eqnarray}
\Pi^{\mu \nu}_\mathrm{CFJ} = -\frac{e^2 m^2}{4 \pi ^2}\left(3 b^\alpha-2 a_F^{\alpha}\right)k_\beta{\epsilon_\alpha}^{\beta\mu\nu},
\end{eqnarray}
where we have assumed our third-rank constant tensors to look like
\begin{subequations}\label{baF}
\begin{eqnarray}
b^{(5)\mu \alpha \beta} &=& b^{\mu}g^{\alpha\beta}+b^{\alpha}g^{\mu\beta}+b^{\beta}g^{\mu\alpha}, \\
a_F^{(5)\mu \alpha \beta} &=& {\epsilon^{\mu\alpha\beta}}_\gamma\, a_F^{\gamma}.
\end{eqnarray}
\end{subequations}
The explanation for this choice, which we use in the next section, looks like follows. While, by definition \cite{KosLi}, the only restriction on the coefficients $b^{(5)\mu \alpha \beta}$ and $a_F^{(5)\mu \alpha \beta}$ is their symmetry or antisymmetry correspondingly with respect to the second and third indices, we require these coefficients to be completely characterized by a single axial vector, first, in order to obtain the CFJ term known to be completely described by such a vector, second, for the sake of simplicity. In principle, this calculation, and studies of other perturbative corrections arising in our theory, can be done for an arbitrary form of these coefficients as well.

We see that our result seems to be ambiguous. Nevertheless, more studies are needed concerning this issue. As we will see further, no CFJ term is generated in another framework, using dimensional regularization and definite symmetrizations (\ref{Rpp}).

\section{Feynman parametrization}\label{fp}

So, as we saw above, in the dimensional regularization scheme, no contribution was generated for $\Pi^{\mu\nu}=\Pi_2^{\mu\nu}+\Pi_3^{\mu\nu}+\Pi_4^{\mu\nu}+\Pi_5^{\mu\nu}+\Pi_6^{\mu\nu}$ in minimal order of $\mathcal{O} (k^2/m^2)$. Thus, let us now consider the higher-order terms of equations~(\ref{foc}) by employing, instead of taking into account only the first term in the derivative expansion of propagators, the Feynman parametrization and dimensional regularization for the expressions involving exact propagators. It is easy to observe that, by power counting, potential divergences can appear in these higher-order contributions, particularly in the higher-derivative CFJ term.

In this way, after considering the splitting 
\begin{subequations}
\begin{eqnarray}
\Pi_2^{\mu\nu} &=& \Pi_{2a}^{\mu\nu}+\Pi_{2b}^{\mu\nu}, \\
\Pi_3^{\mu\nu} &=& \Pi_{3a}^{\mu\nu}+\Pi_{3b}^{\mu\nu}, \\
\Pi_4^{\mu\nu} &=& \Pi_{4a,1}^{\mu\nu}+\Pi_{4a,2}^{\mu\nu}+\Pi_{4b,1}^{\mu\nu}+\Pi_{4b,2}^{\mu\nu}, \\
\Pi_5^{\mu\nu} &=& \Pi_{5a_F}^{\mu\nu}+\Pi_{5b_F}^{\mu\nu}, \\
\Pi_6^{\mu\nu} &=& \Pi_{6a_F}^{\mu\nu}+\Pi_{6b_F}^{\mu\nu},
\end{eqnarray}
\end{subequations}
introducing the Feynman parameter $x$, and calculating the trace, for the coefficient $b^{(5)\alpha\kappa\lambda}$, we obtain
\begin{subequations}\label{FPI1}
\begin{eqnarray}
\Pi^{\mu \nu}_{2b} &=& 4ie^2 \mu^{4-D} \int_0^1 dx\int \frac{d^Dp}{(2\pi)^D} \frac{{g_\lambda}^\mu \left(k_{\kappa }+2 q_{\kappa }\right)}{(p^2-M^2)^2} b^{(5)\alpha\kappa\lambda} {\epsilon_\alpha}^{\nu\sigma\tau}k_\sigma q_\tau, \\
\Pi^{\mu \nu}_{3b} &=& 4ie^2 \mu^{4-D} \int_0^1 dx \int \frac{d^Dp}{(2\pi)^D} \frac{{g_\lambda}^\nu \left(3 k_{\kappa }-2 q_{\kappa }\right)}{(p^2-M^2)^2} b^{(5)\alpha\kappa\lambda} {\epsilon_\alpha}^{\mu\sigma\tau}k_\sigma q_\tau, \\
\Pi^{\mu \nu}_{4b,1} &=& -4ie^2 \mu^{4-D} \int_0^1 dx\ 2x\int \frac{d^Dp}{(2\pi)^D} \frac{q_{\kappa } q_{\lambda }}{(p^2-M^2)^3} b^{(5)\alpha\kappa\lambda} (\left(m^2-q^2\right) {\epsilon_\alpha}^{\mu\nu\sigma}k_\sigma \nonumber\\
&&+\left(2 k\cdot q+m^2-q^2\right) {\epsilon_\alpha}^{\mu\nu\sigma}q_\sigma-2 q^{\nu } {\epsilon_\alpha}^{\mu\sigma\tau}k_\sigma q_\tau+2 q^{\mu } {\epsilon_\alpha}^{\nu\sigma\tau}k_\sigma q_\tau), \\
\Pi^{\mu \nu}_{4b,2} &=& 4ie^2 \mu^{4-D} \int_0^1 dx\ 2(x-1) \int \frac{d^Dp}{(2\pi)^D} \frac{\left(k_{\kappa }-q_{\kappa }\right) \left(k_{\lambda }-q_{\lambda }\right)}{(p^2-M^2)^3} b^{(5)\alpha\kappa\lambda} \left(-\left(k^2+m^2-q^2\right) {\epsilon_\alpha}^{\mu\nu\sigma}q_\sigma \right. \nonumber\\
&&\left.+2 \left(\left(k^{\nu }-q^{\nu }\right) {\epsilon_\alpha}^{\mu\sigma\tau}+\left(q^{\mu }-k^{\mu }\right) {\epsilon_\alpha}^{\nu\sigma\tau}\right)k_\sigma q_\tau+2 \left(k\cdot q+m^2-q^2\right) {\epsilon_\alpha}^{\mu\nu\sigma}k_\sigma\right),
\end{eqnarray}
\end{subequations}
and, for the coefficient $a_F^{(5)\alpha\kappa\lambda}$, we get
\begin{subequations}\label{FPI2}
\begin{eqnarray}
\Pi^{\mu \nu}_{5a_F} &=& -8 i e^2 \mu^{4-D} \int_0^1 dx\int \frac{d^Dp}{(2\pi)^D} \frac{k_{\kappa } {g_\lambda}^\mu}{(p^2-M^2)^2} \nonumber\\
&&\times\ a_F^{(5)\alpha\kappa\lambda} \left({g_\alpha}^\nu \left(k\cdot q+m^2-q^2\right)-q_{\alpha } \left(k^{\nu }-2 p^{\nu }\right)-k_{\alpha } q^{\nu }\right), \\
\Pi^{\mu \nu}_{6a_F} &=& 8 i e^2 \mu^{4-D} \int_0^1 dx\int \frac{d^Dp}{(2\pi)^D} \frac{k_{\kappa } {g_\lambda}^\nu}{(p^2-M^2)^2}  \nonumber\\ 
&&\times\ a_F^{(5)\alpha\kappa\lambda} \left({g_\alpha}^\mu \left(k\cdot q+m^2-q^2\right)-q_{\alpha } \left(k^{\mu }-2 q^{\mu }\right)-k_{\alpha } q^{\mu }\right),
\end{eqnarray}
\end{subequations}
where $q_\mu = p_\mu+(1-x) k_\mu$ is the shifted internal momentum and $M^2=m^2+x(x-1)k^2$. Regarding the coefficient $b_F^{(5)\alpha\kappa\mu}$, after calculating the trace, $\Pi_{5a_F}^{\mu\nu}$ and $\Pi_{6a_F}^{\mu\nu}$ display results involving only odd orders in the internal momentum $p_\mu$. Therefore, the integrals over momenta and corresponding contributions to the effective action are equal to zero, i.e., $\Pi_{a_F}^{\mu\nu}=\Pi_{5a_F}^{\mu\nu}+\Pi_{6a_F}^{\mu\nu}=0$. This is one more argument in favor of vanishing the CFJ-like contributions involving $b_F^{(5)\alpha\kappa\mu}$, which we already noted above. The results for the coefficient $a^{(5)\alpha\kappa\mu}$ will be discussed below.

Now, defining $\Pi^{\mu \nu}_b=\Pi^{\mu \nu}_{2b}+\Pi^{\mu \nu}_{3b}+\Pi^{\mu \nu}_{4b,1}+\Pi^{\mu \nu}_{4b,2}$ and $\Pi^{\mu \nu}_{a_F}=\Pi^{\mu \nu}_{5a_F}+\Pi^{\mu \nu}_{6a_F}$, using Eq.~(\ref{Rpp}), and singling out the divergent terms of the equations~(\ref{FPI1}) and (\ref{FPI2}), we arrive at
\begin{eqnarray}\label{FPI3}
\Pi^{\mu \nu}_b &=& -\frac{e^2 m^2}{4 \pi ^2 \epsilon'} b^{(5)\alpha\kappa\lambda} ({g_\kappa}^\mu k^\beta{\epsilon^\nu}_{\lambda\alpha\beta }-{g_\lambda}^\mu  k^\beta{\epsilon^\nu}_{\kappa\alpha\beta}+{g_\lambda}^\nu k^\beta {\epsilon ^\mu}_{ \kappa \alpha \beta}-{g_\kappa}^\nu k^\beta{\epsilon^\mu}_{ \lambda \alpha \beta}) \nonumber\\
&& +\frac{e^2 k^2}{24 \pi ^2\epsilon'} b^{(5)\alpha\kappa\lambda} (2 g_{\kappa  \lambda}  k^\beta {\epsilon^{ \mu \nu}}_{\alpha\beta}+2 k_{\lambda } {\epsilon ^{\mu \nu}}_{\alpha\kappa}+2 k_{\kappa } {\epsilon ^{ \mu \nu}}_{\alpha \lambda }+{g_\kappa}^\mu k_\beta {\epsilon^{ \nu   \beta}}_{\lambda\alpha} \nonumber\\
&& +{g_\lambda}^\nu k_\beta {\epsilon^{\mu \beta}}_ {\kappa  \alpha} -{g_\kappa}^\nu k_\beta {\epsilon ^{\mu \beta}}_{\lambda\alpha }-{g_\lambda}^ \mu k_\beta {\epsilon ^{\nu\beta}}_{ \kappa \alpha  }) -\frac{e^2}{12 \pi ^2\epsilon'} b^{(5)\alpha\kappa\lambda}  (2 k_{\kappa } k_{\lambda } k^\beta {\epsilon ^{\mu \nu }}_{\alpha \beta} \nonumber\\
&& -k^\nu (k_{\lambda }k_\beta {\epsilon ^{\mu \beta}} _{\kappa \alpha }+k_{\kappa }k_\beta {\epsilon ^{\mu \beta}}  _{\lambda \alpha })+k^\mu (k_{\lambda }k_\beta {\epsilon ^{\nu\beta}}_{\kappa\alpha }+k_{\kappa }k_\beta {\epsilon ^{\nu\beta }}_{\lambda\alpha})) +\text{finite terms}
\end{eqnarray}
and 
\begin{eqnarray}\label{FPI3}
\Pi^{\mu \nu}_{a_F} &=& -\frac{e^2}{3 \pi ^2 \epsilon'} a_F^{(5)\alpha\kappa\lambda} k_{\kappa } \left({g_\lambda}^\mu k_{\alpha } k^\nu-k^2 {g_\alpha}^\nu {g_\lambda}^\mu +{g_\lambda}^\nu \left(k^2 {g_\alpha}^\mu-k_{\alpha } k^\mu \right)\right) +\text{finite terms},
\end{eqnarray}
where $\frac1{\epsilon'}=\frac1{\epsilon}-\ln\frac{m}{\mu'}$, with $\epsilon=4-D$ and $\mu'^2=4\pi\mu^2e^{-\gamma}$. Then, considering the decompositions (\ref{baF}) for the coefficients $b^{(5)\alpha\kappa\lambda}$ and $a_F^{(5)\alpha\kappa\lambda}$, we obtain
\begin{equation}
\Pi^{\mu \nu}_{b}+\Pi^{\mu \nu}_{a_F} = -\frac{ e^2 k^2}{3 \pi ^2\epsilon'} (2a_F^\alpha -b^\alpha) k_\beta {\epsilon_\alpha} ^{\beta\mu\nu} +\text{finite terms},
\end{equation}
which has the tensorial structure of a higher-derivative CFJ term \cite{Leite:2013pca}. Finally, in order to eliminate these divergent terms, we can consider, e.g., $a_F^\kappa=\frac12b^\kappa$.

With this choice, the finite contribution assumes the form
\begin{eqnarray}\label{CS}	
\Pi^{\mu \nu}_{b}+\Pi^{\mu \nu}_{a_F} = \frac{e^2}{12\pi^2} \left[k^2 +6 m^2 -\frac{24m^4}{\sqrt{k^2 \left(4 m^2-k^2\right)}} \cot^{-1}\left(\sqrt{\frac{4 m^2}{k^2}-1}\right)\right] b_\alpha k_\beta \epsilon^{\alpha\beta\mu\nu}.
\end{eqnarray}
We can easily verify the derivative expansion result (\ref{O0}) by taking the limit $k^2 \ll m^2$ $(m\neq 0)$ in the above equation~(\ref{CS}), i.e., we get
\begin{eqnarray}
\Pi^{\mu \nu}_{b}+\Pi^{\mu \nu}_{a_F} =  -\frac{e^2}{60\pi^2m^2} k^4 b_\alpha k_\beta \epsilon^{\alpha\beta\mu\nu} +{\cal O}\left(\frac{k^6}{m^6}\right),
\end{eqnarray}
which means that the corresponding CFJ term is zero, as expected. Note that the higher-derivative CFJ term is zero as well. Interestingly, in this case, the dimension-5 LV operators generate the higher-derivative terms beginning from five derivatives, while lower-derivative terms vanish. This situation can be compared with that in \cite{MNP}, where the dimension-3 LV operator generates the higher-derivative terms beginning from three derivatives while the one-derivative term vanishes within the dimensional regularization scheme.

Now, let us discuss the results for the coefficient $a^{(5)\alpha\kappa\lambda}$. After calculating the trace and introducing the Feynman parameter $x$, we obtain
\begin{subequations}\label{FPI}
\begin{eqnarray}
\Pi^{\mu \nu}_{2a} &=& -4e^2 \mu^{4-D} \int_0^1 dx\int \frac{d^Dp}{(2\pi)^D} \frac{{g_\lambda}^\mu \left(k_{\kappa }+2 q_{\kappa }\right)}{(p^2-M^2)^2} \nonumber\\
&&\times\ a^{(5)\alpha\kappa\lambda}\left({g_\alpha}^\nu \left(k\cdot q+m^2-q^2\right)-q_{\alpha } \left(k^{\nu }-2 q^{\nu }\right)-k_{\alpha } q^{\nu }\right), \\
\Pi^{\mu \nu}_{3a} &=& 4e^2 \mu^{4-D} \int_0^1 dx\int \frac{d^Dp}{(2\pi)^D} \frac{{g_\lambda}^\nu \left(3 k_{\kappa }-2 q_{\kappa }\right)}{(p^2-M^2)^2}  \nonumber\\
&&\times\ a^{(5)\alpha\kappa\lambda} ({g_\alpha}^\mu \left(k\cdot q+m^2-q^2\right)-q_{\alpha } \left(k^{\mu }-2 q^{\mu }\right)-k_{\alpha } q^{\mu }), \\
\Pi^{\mu \nu}_{4a,1} &=& 4e^2 \mu^{4-D} \int_0^1 dx\ 2x\int \frac{d^Dp}{(2\pi)^D} \frac{q_{\kappa } q_{\lambda }}{(p^2-M^2)^3} a^{(5)\alpha\kappa\lambda} \left(k_{\alpha } \left(m^2-q^2\right) g^{\mu  \nu }- \left({g_\alpha}^\mu \left(k^{\nu }-q^{\nu }\right) \right.\right. \\
&&+ \left.\left.{g_\alpha}^\nu \left(k^{\mu }-q^{\mu }\right)\right) \left(m^2-q^2\right)+q_{\alpha } \left(g^{\mu  \nu } \left(2 k\cdot q+m^2-q^2\right)-2 k^{\nu } q^{\mu }-2 q^{\nu } \left(k^{\mu }-2 q^{\mu }\right)\right)\right), \nonumber\\
\Pi^{\mu \nu}_{4a,2} &=& 4e^2 \mu^{4-D} \int_0^1 dx\ 2(x-1) \int \frac{d^Dp}{(2\pi)^D} \frac{\left(k_{\kappa }-q_{\kappa }\right) \left(k_{\lambda }-q_{\lambda }\right)}{(p^2-M^2)^3} a^{(5)\alpha\kappa\lambda} \left(\left(q^{\nu } {g_\alpha}^\mu+q^{\mu } {g_\alpha}^\nu \right) \right. \nonumber\\
&& \times\left.\left(-k^2+2 k\cdot q+m^2-q^2\right) -2 k_{\alpha } \left(g^{\mu  \nu } \left(k\cdot q+m^2-q^2\right)-k^{\nu } q^{\mu }-q^{\nu } \left(\left(k^{\mu }-2 q^{\mu }\right)\right)\right) \right. \nonumber\\
&& +\left.q_{\alpha } \left(g^{\mu  \nu } \left(k^2+m^2-q^2\right) -2 k^{\nu } q^{\mu }-2 q^{\nu } \left(k^{\mu }-2 q^{\mu }\right)\right)\right).
\end{eqnarray}
\end{subequations}
with again $q_\mu = p_\mu+(1-x) k_\mu$ and $M^2=m^2+x(x-1)k^2$. Thus, by setting $\Pi^{\mu \nu}_a=\Pi^{\mu \nu}_{2a}+\Pi^{\mu \nu}_{3a}+\Pi^{\mu \nu}_{4a,1}+\Pi^{\mu \nu}_{4a,2}$, we can write $\Pi^{\mu \nu}_a$ as
\begin{equation}
\Pi^{\mu \nu}_{a} = \frac{ i e^2}{3 \pi ^2\epsilon'} a^{(5)\alpha\kappa\lambda} k_{\kappa } ({g_\lambda}^{\mu } k_{\alpha } k^\nu-k^2 {g_\alpha}^\nu {g_\lambda}^\mu+{g_\lambda}^\nu (k^2 {g_\alpha}^\mu-k_{\alpha } k^\mu)) +\text{finite terms},
\end{equation}
where we have singled out the divergent part. This expression vanishes if we assume $a^{(5)\alpha\kappa\lambda}$ to be symmetric in two last indices, e.g., by considering
\begin{equation}
a^{(5)\alpha\kappa\lambda} = a^{\kappa}g^{\alpha\lambda}+a^{\lambda}g^{\alpha\kappa}.
\end{equation}
With this choice, the finite part vanishes as well, so, $\Pi^{\mu \nu}_{a}=0$.

Therefore, rewriting $\Pi^{\mu \nu}=\Pi^{\mu \nu}_{a}+\Pi^{\mu \nu}_{b}+\Pi^{\mu \nu}_{a_F}+\Pi^{\mu \nu}_{b_F}$, we have obtained the higher-derivative contributions to the effective action in the CPT-odd sector (\ref{1}), given by
\begin{eqnarray}\label{hdCS}	
\Pi^{\mu \nu} = \frac{e^2}{12\pi^2} \left[k^2 +6 m^2 -\frac{24m^4}{\sqrt{k^2 \left(4 m^2-k^2\right)}} \cot^{-1}\left(\sqrt{\frac{4 m^2}{k^2}-1}\right)\right] b_\alpha k_\beta \epsilon^{\alpha\beta\mu\nu}.
\end{eqnarray}

Considering behaviour of this result in the $k\to 0$ limit, we conclude that the CFJ term is not generated, which agrees with the result (\ref{O0}) obtained within the derivative expansion approach. 
It is also worth commenting that the divergent contributions can be canceled by imposing a given proportionality between the coefficients $a_F^\mu$ and $b^\mu$.

\section{Summary}\label{summ}

 Let us discuss our results. In this paper, we considered the LV extended QED involving all nonminimal CPT-odd dimension-5 couplings, whose perturbative impacts have not been ever considered. For this theory, we studied the possibility of perturbative generation of the LV terms, among them the CFJ term. We performed the explicit calculations within two frameworks, namely, the derivative expansion and Feynman parametrization approaches. 
We explicitly demonstrated that no generation of the term CFJ occurs when we consider the dimensional regularization scheme, which is consistent with claims made in \cite{Altschul:2019eip,Karki:2020rgi}.


One important conclusion of our paper is the confirmation that LV theories with nonminimal (nonrenormalizable) couplings can yield finite results. A similar situation was earlier shown to take place for the magnetic coupling generating finite aether and three-derivative terms \cite{aether,aether1,MNP}. Another important conclusion is that we found one more manner allowing to generate finite higher-derivative CPT-odd terms, based on coupling different from that one used in \cite{MNP}.


 Natural extensions and continuations of our paper can be developed. First, it is natural to study higher orders in derivative expansions of the effective action in the nonminimal LV QED, especially, it is interesting to generalize our results through obtaining the aether term with the use of the dimension-5 coupling considered in this paper applying various prescriptions, and check whether this term, for certain couplings, could be finite and ambiguous, or even vanish in some of the regularization schemes. Second, it is natural to investigate perturbative impacts of other dimensions-5 and 6 LV operators, which do not yield the CFJ contribution and therefore were not treated in this paper, and thus, extend the tables relating couplings and new terms in the gauge sector generated with their use, presented in \cite{ourrev}. We plan to pursue these aims in our next papers.

{\bf Acknowledgments.}  This work was partially supported by Conselho
Nacional de Desenvolvimento Cient\'{\i}fico e Tecnol\'{o}gico (CNPq). The work by A. Yu. P. has been supported by the
CNPq (project No. 301562/2019-9) and by R. M. -- by the CAPES/FAPEAL (project No. 60030-001204/2018).

\end{document}